\documentclass[conference]{IEEEtran}
\IEEEoverridecommandlockouts
\usepackage{multirow}
\usepackage{cite}
\usepackage{amsmath,amssymb,amsfonts}
\usepackage{algorithmic}
\usepackage{graphicx}
\usepackage{textcomp}
\usepackage{xcolor}
\usepackage[a4paper, total={184mm,239mm}]{geometry}
\def\BibTeX{{\rm B\kern-.05em{\sc i\kern-.025em b}\kern-.08em
    T\kern-.1667em\lower.7ex\hbox{E}\kern-.125emX}}

\usepackage{bm}
\usepackage{subcaption}

\begin{document}

\title{\LARGE{Stealthy Attack on Algorithmic-Protected DNNs via Smart Bit Flipping}}

\author{\IEEEauthorblockN{Behnam Ghavami, Seyd Movi, Zhenman Fang, Lesley Shannon}
\IEEEauthorblockA{Simon~Fraser~University, Burnaby, BC, Canada\\
Emails: \{behnam\_ghavami, zhenman, lesley\_shannon\}@sfu.ca}}

\maketitle
\begin{abstract}
Recently, deep neural networks (DNNs) have been deployed in safety-critical systems such as autonomous vehicles and medical devices. 
Shortly after that, the vulnerability of DNNs were revealed by stealthy adversarial examples where crafted inputs---by adding tiny perturbations to original inputs---can lead a DNN to generate misclassification outputs.
To improve the robustness of DNNs, some algorithmic-based countermeasures against adversarial examples have been introduced thereafter.

In this paper, we propose a new type of stealthy attack on protected DNNs to circumvent the algorithmic defenses: via smart bit flipping in DNN weights, we can reserve the classification accuracy for clean inputs but misclassify crafted inputs even with algorithmic countermeasures. 
To fool protected DNNs in a stealthy way, we introduce a novel method to efficiently find their most vulnerable weights and flip those bits in hardware. Experimental results show that we can successfully apply our stealthy attack against state-of-the-art algorithmic-protected DNNs.

\end{abstract}



\section{Introduction}\label{sec:intro}
In the past few years, DNNs have achieved an amazing success in many areas, especially in computer vision and speech recognition.
With respect to their great performance and autonomous nature, DNNs have been recently deployed in critical systems such as personal identity recognition systems, self-driving cars, aircraft control and medical devices \cite{shafique2020robust}.
Therefore, it is very important to study the vulnerability and safeguard of such DNN-based systems under various attacks. 

\begin{figure}[!tb]
	\begin{center}
		\includegraphics[width = 0.5\textwidth]{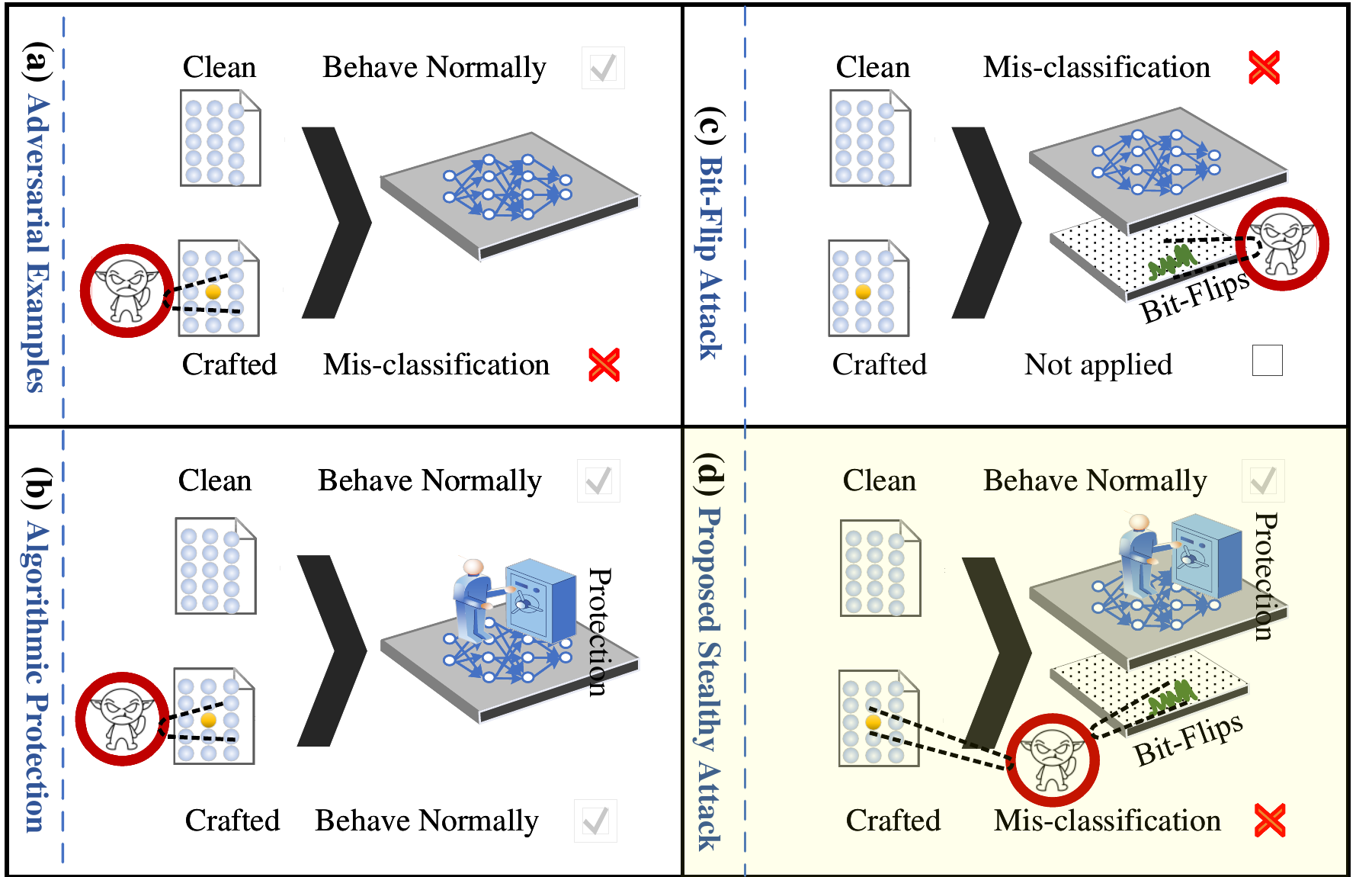}
		\vspace{-0.1in}
		\caption{Our ``stealthy'' attack against a protected DNN via smart bit flipping of DNN weights. Both the DNN inputs and parameters are perturbed. The DNN works correctly for clean inputs, but misclassifies inconspicuous crafted inputs.}
		\vspace{-0.1in}
		\label{fig:CLAsDNNMODEL}
	\end{center}	
\vspace{-0.1in}
\end{figure}

Recently, DNNs were found vulnerable to adversarial example attacks \cite{szegedy2013intriguing,goodfellow2014explaining,moosavi2016deepfool,carlini2017towards,moosavi2017universal,hayes2018learning,yuan2019adversarial,chaubey2020universal}, which fool DNNs to misclassify crafted inputs with imperceptible perturbations (shown in Figure~\ref{fig:CLAsDNNMODEL}(a)).
Such attacks are stealthy and hard to notice for a user who only has access to original clean inputs~\cite{chaubey2020universal}. Thus, they are more harmful and have raised serious concerns.

To improve the robustness of DNN models, various algorithmic defenses have been introduced thereafter \cite{chakraborty2018adversarial}.
The state-of-the-art adversarial defenses train the DNNs using both clean and adversarial examples \cite{sinha2017certifying,madry2017towards,zhang2019theoretically,wong2020fast}. As a result, when confronted with adversarial examples, such trained DNN models would behave more robust (shown in Figure~\ref{fig:CLAsDNNMODEL}(b)).

In this paper, we introduce a new ``stealthy bit-flip attack'' against algorithmic-protected DNNs to circumvent algorithmic countermeasures. 
This is based on our observation that flipping the bits of DNN hardware parameters---i.e., DNN weights in this paper---can cause the robustness and accuracy of a DNN changing in different ways.
Therefore, we propose an adversarial perturbation attack on DNN weights such that the robustness of a DNN has a significant drop but its accuracy remains almost the same to ensure the stealthiness. 

It is worth noting that while conventional bit-flip attacks \cite{liu2017fault, breier2018practical, rakin2019bit, yao2020deephammer, dumont2021overview, zhao2019fault, rakin2020t, bai2021targeted} aim to degrade the overall accuracy of a DNN (for clean inputs, shown in Figure~\ref{fig:CLAsDNNMODEL}(c)), our attack (shown in Figure~\ref{fig:CLAsDNNMODEL}(d)) targets misclassification for selectively crafted inputs (i.e., degrading the robustness) by the protected DNNs with algorithmic defenses, while maintaining a similar accuracy for clean inputs.
As a result, a user of a DNN system (with algorithmic defenses) would be unaware of the threat when such an attack occurs, which can cause higher calamity. 
For example, applying this attack (i.e., injecting specific bit-flips to DNN weights via software) against a face identity recognition system could cause a person's face---to which an imperceptible perturbation (e.g., a special wearing or masking) is attached---to be misclassified as another person, while keeping the classification of all normal faces accurate. 

We formulate our new stealthy attack as a mathematical optimization problem: through smart bit flipping of the weights, we aim to reduce the DNN robustness under the constraint that its accuracy loss is below a perceptible threshold. 
To find the most vulnerable bits, we introduce an iterative gradient-based bit search algorithm. 
Simulation results confirm that our attack can successfully circumvent the state-of-the-art algorithmic countermeasure \cite{zhang2019theoretically}: flipping a small set of weight bits (e.g., 30 out of 493,648 bits in LeNet-3) can result in a significant robustness drop (e.g., 59.9\% in  LeNet-3) while there is negligible accuracy loss.

\section{Related Work}
\label{sec:related}


In this section, we first briefly review related attacks against DNNs, which could be classified as attacks by changing the DNN models, inputs, or parameters. Then we summarize the novelty of our work.


\subsection{Adversarial Trojan Insertion into DNN Models} \label{subsec:trojan-attack}
Attackers may insert various types of malicious components (i.e., Trojans) into DNN models. Typically, Trojan requires hacking the training flow to be triggered via a specific input pattern, which can lead to DNN misclassification \cite{liu2017trojaning}. Li et al. \cite{li2018hu} introduced a hardware Trojan circuit to implement malicious DNN models. 
Clements et al. \cite{clements2018hardware} exploited the multiplexer logic to alter the internal structure of certain operations to inject malicious behavior.
Zhao et al. \cite{zhao2019memory} introduced a memory Trojan attack towards DNN accelerator platforms without toolchain manipulation. 
A recent more advanced Trojan attack, Targeted Bit Trojan (TBT), leverages flipping bits of weights of last-layer neurons to trigger Trojan with no need for supply chain access \cite{rakin2020tbt}.

\subsection{Adversarial Parameter Perturbation} \label{subsec:parameter-attack}
Lately, some researchers study the vulnerability of DNNs against adversarial parameter perturbation (such as weights and biases) using 
intentional memory fault injections. These attacks are divided into two categories.

\subsubsection{Untargeted Bit-flip Attack} \label{subsubsec:untargeted}
The main focus of untargeted bit-flip attack is reducing the overall prediction accuracy of the DNN classifier to be as low as random guess.
Liu et al. \cite{liu2017fault} were the first to explore memory fault injection of a DNN hardware to achieve misclassification. 
Breier et al. \cite{breier2018practical} experimentally showed what types of memory fault attacks are achievable in practice. 
Rakin et al. \cite{rakin2019bit} presented a method to find the specific memory fault patterns that can cause important destruction to the DNN accuracy.
Taking advantage of the well-known row hammer attack \cite{kim2014flipping}, Yao et al. \cite{yao2020deephammer} attempted to attack a DNN hardware where the network weights are stored in DRAM.
Dumont et. al \cite{dumont2021overview} presented how laser injection with state-of-the-art equipment threats against the DNN inference.

\subsubsection{Targeted Bit-flip Attack} \label{subsubsec:targeted}
Targeted adversarial attacks pose a greater threat as they give the attacker precise control on the malicious behavior.
Lately, Zhao et al. \cite{zhao2019fault} introduced a bit flipping attack on a DNN classifier in order to stealthily misclassify a few predefined inputs. However, it is hard for this method to adapt properly to (mis)classify previously unseen inputs. Hence, it could not be applied in more general scenarios such as autonomous vehicles, where the \textit{domain generalization} is important as well.
Recent works can perform a targeted bit flip attack on full precision models \cite{zhao2019fault}, although they require large amounts of weight perturbation.
Rakin et al. \cite{rakin2020t} introduced an adversarial bit flip attack on quantized DNN models where the main goal is to identify the weights that are highly associated with the misclassification of a targeted output. Lately, Bai et al. \cite{bai2021targeted} formulated the targeted adversarial bit flipping as a binary integer programming.


\subsection{Adversarial Input Perturbation} \label{subsec:input-attack}
DNNs are also susceptible to projected small input perturbations. Adversarial examples can mislead state-of-the-art DNN classifiers to make erroneous predictions \cite{yuan2019adversarial}. The size of the perturbation is at the heart of the adversarial example attack which enables the possibility of the stealthy threats; in fact, such inputs are built on the premise of a small perturbation. The attacker wants the perturbed input to be as close to the original input as possible when designing an adversarial example. When it comes to images, it is close enough that a human observer could hardly detect the perturbation. Szegedy et al. \cite{szegedy2013intriguing} first revealed that DNNs are vulnerable to such stealthy adversarial inputs. Recently, plenty of attacks based on the adversarial perturbations of DNN inputs have been introduced \cite{chakraborty2018adversarial}. We classify these attacks into two main categories.

\subsubsection{Per-Instance Model}
In this category, 
the type of augmented adversarial inputs highly depends on the input images \cite{moosavi2016deepfool}.
Some techniques focus on maximizing the loss function of the target model by changing the input in the opposite direction of its gradients \cite{goodfellow2014explaining}.
Other methods such as \cite{carlini2017towards} take advantage of an objective function to alter the input image that may cause output misclassification.

\subsubsection{Universal Model}
Universal perturbation attacks exploit image-agnostic perturbations to misclassify the identity of an object to be selected later in the field.
Moosavi et al. \cite{moosavi2017universal} first introduced a universal adversarial perturbation which could fool DNN models.
Recently, researchers introduce various methods to extend the original universal adversarial attack \cite{moosavi2017universal}.
Some methods \cite{mopuri2017fast}\cite{mopuri2018generalizable} exploited a data-independent approach to generate adversarial perturbation vector by modifying the features extracted at various layers of the network.
Hayes et al. \cite{hayes2018learning} generated adversarial perturbation inputs with the generative adversarial networks.
Several other studies have also lately introduced other methods to create adversarial attacks \cite{chaubey2020universal}.


\subsection{Novelty of Our Work}\label{subsec:nov}
To the best of our knowledge, we are the first to propose a stealthy bit-flip attack over protected DNNs with algorithmic countermeasures.
This new type of attack has a main objective: \textit{All the benign (clean) inputs are classified into correct categories and only crafted (selected) inputs are misclassified, and therefore, \textbf{the attack cannot be easily detected via user inspection or screening, which makes it more stealthy.}} 

In comparison to stealthy adversarial example attacks in Section \ref{subsec:input-attack}, our goal is attack against \textbf{protected DNNs}.
Although prior bit-flip attacks can be deployed against protected DNNs, all untargeted ones, discussed in Section \ref{subsubsec:untargeted}, try to drop the DNN accuracy, which are not \textbf{stealthy}. Moreover, compared with targeted bit-flip attacks in Section \ref{subsubsec:targeted}, which always misclassifiy samples from one source class, it is difficult to conduct a \textbf{screening test} for our proposed attack because it can be used selectively for one and more source types of unseen inputs. 
In contrast to a more closely related work, i.e., TBT attack in Section~\ref{subsec:trojan-attack}, which can target a Trojan through bit-flipping and input modification as a trigger, our attack needs to modify inputs with imperceptible perturbations that are not visible to human eyes because our input perturbation is based on adversarial examples.
However, in order to hide the input trigger from human eyes, TBT sacrifices its strength which may \textbf{jeopardise its stealthiness}.

\section{Assumptions of Our Proposed Stealthy Attack}
\label{sec:assumption}
In this section, we present the protected DNN model with algorithmic countermeasures and our attack assumptions.

\subsection{Protected DNN Model with Adversarial Training}\label{subsec:pdnn}
A DNN model usually consists of multiple layers of neurons and neurons in adjacent layers are connected by weighted edges. The weights are optimized in the training stage and are usually stored in the memory of the DNN hardware and remain fixed afterwards during the inference stage.

In this paper, we assume that before deploying a DNN model in the inference stage, an efficient method is in place to protect it against conventional adversarial examples attacks (we call it as protected DNN).
Among various defense strategies, adversarial training currently proves to be the most effective against adversarial attacks \cite{sinha2017certifying,madry2017towards,zhang2019theoretically,wong2020fast}  and it is one of the few defenses that withstands against strong adversarial attacks \cite{bai2021recent}. 

\subsection{Level of Access to DNN at Inference Stage}\label{subsec:access}
We assume that the attacker has enough information about the DNN architecture and particularly its weight parameters, as well as the hardware structure of the deployment platform. Indeed, we use the standard white-box attack threat model assumption in this work, which is consistent with previous DNN bit-flip attacks \cite{zhao2019fault,rakin2020t,bai2021targeted}. Such an assumption is reasonable for the following reasons. First, as the training  process is typically costly, developers tend to utilize pre-trained models released by third parties (e.g., ModelZoo \cite{modelzoosite}) to accelerate the time to market of the system. Second, even with private models, adversaries can learn about model parameters through a variety of information leak via side channels attacks \cite{gongye2020reverse,wei2020leaky,xiang2020open}.


\subsection{Imperceptible Input Modification}
In the same way that conventional adversarial example attacks have been used \cite{moosavi2016deepfool,carlini2017towards,moosavi2017universal,hayes2018learning,yuan2019adversarial,chaubey2020universal}, our attack requires modifying the input (images) used by the DNN. We assume that the adversary could potentially attempt to imperceptibly manipulate the input images. Some research works showed the feasibility of such modification for a state-of-the-art vision classifier \cite{kurakin2016adversarial} and face recognition model \cite{sharif2016accessorize}. For example, physical adversarial traffic signs could be created by maliciously altering the sign itself, such as with stickers or paint.

\subsection{Precise Memory Fault Injection}
Similar to how traditional bit-flip attacks have been used \cite{zhao2019fault,rakin2020t,bai2021targeted}, we also assume that the adversary will attempt to precisely inject a small amount of faults into the computing memory that is deployed for DNN inference. For example, rowhammer attack \cite{kim2014flipping}, 
can inject faults into a computer main memory. Rowhammer is particularly amenable to practical real-world exploitation, including browsers, mobile and servers, as it is the common instance of \textbf{software-induced bit-flips attacks}. Precise surgical rowhammering \cite{tatar2018defeating} can be used as it has been shown to be the most effective in inducing specific bit flips at the targeted locations.


\section{Smart Bit Flipping on Protected DNNs}
\label{sec:proposed}

In this section, we present our  stealthy attack over protected DNNs, making adversarial attacks to be almost as easy as before algorithmic protection.
Our goal is to figure out how to flip certain DNN weight bits in a way such that it significantly reduces the robustness of the protected DNN while causing negligible accuracy loss. First, we define the security threat model and mathematically formulate our attack as an optimization problem. Then, to find the desired bit flip candidates, we propose an iterative gradient-based bit search algorithm.

\subsection{Definition of Threat Model}\label{subsec:threat-model}

\subsubsection{Terminology} $f$ denotes a DNN classifier function, $\bm{x}^{(i)}$ denotes the $i^{th}$ input image and $y^{(i)}$ denotes its associated true class label. $f(\bm{x}^{(i)})$ denotes the output probability vector of all predicted class labels (with confidence values) for the $i^{th}$ input image. $k(\bm{x}^{(i)})$ denotes the predicted class label (with the highest confidence) of classifier $f$ for input $\bm{x}^{(i)}$.
\subsubsection{Adversarial Accuracy Loss Function}
The \textit{loss function} is a metric to evaluate how accurate the DNN model predicts for a given input compared to the true label.
Formally, the loss value per input $(\bm{{x}^{(i)}},y^{(i)})$ can be defined as:
\begin{equation}\label{eq1}
    \bm{\mathcal{L}}(\bm{x}^{(i)},y^{(i)})=D(f(\bm{x}^{(i)}),y^{(i)})
\end{equation}
where $D$ is a distance metric that varies for problems. In this paper we use the cross-entropy loss function \cite{goodfellow2016deep}.

With an adversarial attack on the DNN weights $\bm{w}$, we define the \textit{adversarial  loss function} of a DNN model as:
\begin{equation}
    \bm{\mathcal{L}}(\bm{x}^{(i)},y^{(i)};\bm{w})=D(f(\bm{x}^{(i)};\bm{w}),y^{(i)})
\end{equation}
where  $f(\bm{x}^{(i)};\bm{w})$ is
the output probability vector of all predicted class labels (with confidence values) for input $\bm{x}^{(i)}$ under attack of classifier $f$ on its weight parameters $\bm{w}$.

The adversarial accuracy loss function on the entire distribution dataset $\mathbb{D}$ can be defined as:
\begin{equation}
    \bm{\mathcal{L}}(\mathbb{D};\bm{w}) = \mathbb{E}_{(\bm{x},y) \in \mathbb{D}} \bm{\mathcal{L}}(\bm{x},y;\bm{w})
\end{equation}
where $\mathop{\mathbb{E}}_{\bm{x},y}(.)$ denotes the expectation function.
\subsubsection{DNN Model Overall Accuracy}
For a given test dataset $\mathbb{D}$, the overall accuracy of a DNN model (i.e., a classifier $f$) is defined as: 
\begin{equation}\label{Accuracy}
    ACC = \mathbb{E}_{(\bm{x},y) \in D} \bm{1}|\{k(\bm{x},\bm{w}) == y\}
\end{equation}
where $\bm{1}|\{event\}$ represents an indicator function that is $1$ if $event$ happens and $0$ otherwise. $\bm{w}$ notes the weights of the model. We use the overall accuracy to evaluate the stealthiness of our proposed attack.

\subsubsection{Adversarial Robustness}
\label{subsubsec:robustness}

For a well-trained DNN model, its classification accuracy is usually very high in a well controlled setting. However, prior studies show that these models are fascinatingly vulnerable to small perturbations on inputs (i.e., adversarial examples). 
Informally, an adversarial example is an imperceptible perturbation ($\epsilon$) that is added to an input image, which can change the classifier's prediction \cite{goodfellow2014explaining}.

To quantify the robustness of a classifier $f$, we can find the minimum perturbation vector $\bm{r}$ that is sufficient to change the predicted class label $k(\bm{x}^{(i)})$ of the classifier for input $\bm{x}^{(i)}$. According to \cite{moosavi2016deepfool}, formally, the robustness of a classifier $f$ for input $\bm{x^{(i)}}$, denoted as $\Delta(f,\bm{x^{(i)}})$, is defined as:  
\begin{equation}
    \Delta(f,\bm{x^{(i)}}) = min_{\boldsymbol{\bm{r}}} ||\bm{r}||_2  \quad s.t. \quad k(\bm{x^{(i)}}+\bm{r})\neq k(\bm{x^{(i)}}) 
\end{equation}
where $||.||_2$ denotes the Euclidean norm.

The value of $\Delta(f,\bm{x^{(i)}})$ depends on the difference between the probability values of the predicted class with the highest confidence ($c^*$) and the closest one to it ($\hat{c}$). Hence, we have: 
\begin{equation}\label{eq6}
    \Delta(f,\bm{x}) \propto |f_{\hat{c}}(\bm{x};\bm{w})-f_{c^*}(\bm{x};\bm{w})|
\end{equation}
where $f_{c^*}(\bm{x};\bm{w})$ is the probability of the predicted class with the highest confidence and $f_{\hat{c}}(\bm{x};\bm{w})$ is the probability of the closest class with the second highest confidence.

According to \cite{moosavi2016deepfool}, we define the \emph{adversarial robustness} of classifier $f$ on the entire distribution dataset $\mathbb{D}$ with respect to the weights parameter $\bm{w}$ as: 
\begin{equation}\label{eq:robustness}
\begin{split}
    &\rho_{adv}^f(\mathbb{D};\bm{w}) = \mathop{\mathbb{E}}_{\bm{x} \in \mathbb{D}} \frac{|f_{\hat{c}}(\bm{x};\bm{w})-f_{c^*}(\bm{x};\bm{w})|}{||\bm{x}||_2}\\
\end{split}
\end{equation}
where $\rho_{adv}^f(.)$ denotes the adversarial robustness of classifier $f$, $\mathop{\mathbb{E}}_{\bm{x}}(.)$ is the expectation function.

\subsubsection{Stealthy Threat}
In some application domains such as personal identity recognition systems, a ``stealthy" attack on their underlying DNN models becomes one of the most important security concerns \cite{sharif2016accessorize}. One difficulty that attackers face in such application domains is that manipulating parameters to evade the DNN classifiers might be easily observed from outside the systems. For example, attackers can circumvent a DNN access control device used in banks, however, these attackers may draw increased attention from bystanders.

We call a bit-flip attack against a DNN (already equipped with algorithmic defenses in our assumption) as a \textit{stealthy threat} if it could produce unexpected classification behavior on crafted inputs while it does not affect the normal behavior of the classifier on benign (clean) inputs. 

\subsection{Problem Formulation of Stealthy Attack}\label{subsec:formulation}
Based on our observation, changing the weight bits in a DNN model may affect the \emph{adversarial accuracy loss} and \emph{adversarial robustness} in different ways. There is an interesting scenario where weight perturbations could significantly decrease the adversarial robustness while negligibly increasing the adversarial accuracy loss. This provides a ground for our stealthy attack against protected DNNs. 

Mathematically, we formulate our stealthy attack against protected DNNs as an optimization problem. Our goal is to find those vulnerable DNN weight bits to minimize the adversarial robustness, under the constraint that the adversarial accuracy loss is below a perceptible threshold, i.e.,
\begin{equation}\label{optimization}
\begin{split}
    &max_{\bm{\hat{\{B_l\}}}} \quad \rho_{adv}^f(\mathbb{D};\bm{\{B_l\}_{l=1}^L}) -\rho_{adv}^f(\mathbb{D};\bm{\hat{\{B_l\}}_{l=1}^L}) \\
    & s.t. \quad \quad\quad \bm{\mathcal{L}}(\mathbb{D};\bm{\hat{\{B_l\}}_{l=1}^L})-\bm{\mathcal{L}}(\mathbb{D};\bm{\{B_l\}_{l=1}^L})<\delta
\end{split}
\end{equation}
where $\bm{\{B_l\}_{l=1}^L}$ is the original weight bits from layer $1$ to $L$ of a well-trained DNN and $\bm{\hat{\{B_l\}}_{l=1}^L}$ is the modified weight bits. $\delta$ denotes an attacker-defined constraint that is application dependent to satisfy the ``stealthy threat" attribute of the attack.

\subsection{Gradient-based Bit Flip Attack}\label{subsec:bit-search}
We use the mathematical gradient concept to quantify the impact of weight bits on adversarial robustness and adversarial loss as:
\begin{equation}\label{robust-eq}
    \nabla_{\bm{b}} \rho_{adv}^f =[\frac{\partial \rho_{adv}^f}{\partial \bm{b}_{N-1}},\cdots,\frac{\partial \rho_{adv}^f}{\partial \bm{b}_{0}}]
\end{equation}
\begin{equation}\label{loss-eq}
    \nabla_{\bm{b}} \mathcal{L}=[\frac{\partial \mathcal{L}}{\partial \bm{b}_{N-1}},\cdots,\frac{\partial \mathcal{L}}{\partial \bm{b}_{0}}]
\end{equation}
where $\bm{b}$ shows binary representation of a weight using $N$ number of bits.

The main idea is to flip the bits along the direction opposite of the gradient of adversarial robustness and adversarial loss with regard to weight bits. Hence, we define $sign(.)$ to represent the gradient direction, where
$sign(.)\in \{0,1\}$. In order to simultaneously decrease robustness and preserve accuracy, we introduce Table \ref{tab:TruthT} to mathematically express the possibility of all changes in the gradient values and bit flipping. This truth table shows the original bit $\bm{b_i}$, $sign({\partial \rho_{adv}^f}^{}/{\partial \bm{b_i}})$ and $sign({\partial \mathcal{L}}^{}/{\partial \bm{b_i}})$ as the possible states, and $m$ shows whether there should be a flip of original bits. To make a decision about whether a bit should be flipped, we formulate $m$ as:
\begin{equation}\label{mask}
    \bm{m} = \neg \Bigg[\left(\bm{b} \oplus sign \left(\nabla_b \rho_{adv}^f \right)\right) \lor \left(sign \left(\nabla_b \rho_{adv}^f \right) \oplus sign \left(\nabla_b \mathcal{L}\right)\right)\Bigg] 
\end{equation}
where $\oplus$, $\lor$ and $\neg$ present bit wise \emph{xor}, \emph{or} and \emph{not} operators. 

\begin{table}[hbt!]
\centering
\caption{Truth table of bit flip attack. $\bm{b_i}$ is the original bit. $m$ indicates whether there should be a flip of $\bm{b_i}$.}
\vspace{-0.05in}
\label{tab:TruthT}
\begin{tabular}{ |ccc|c| }  
\hline
$\bm{b_i}$ & $sign({\partial \rho_{adv}^f}^{}/{\partial \bm{b_i}})$ & $sign({\partial \mathcal{L}}^{}/{\partial \bm{b_i}})$ & $m$ \\ 
\hline
0 & 0 (-) & 0 (-) &  1\\ 
0 & 0 (-) & 1 (+) & 0\\
0 & 1 (+) & 0 (-) & 0\\
0 & 1 (+) & 1 (+) & 0\\
1 & 0 (-) & 0 (-)  & 0\\ 
1 & 0 (-) & 1 (+) & 0\\
1 & 1 (+) & 0 (-)  & 0\\
1 & 1 (+) & 1 (+)  & 1\\
\hline
\end{tabular}
\vspace{-0.05in}
\end{table}

\subsection{Iterative Bit Search}
The number of weight bits in a DNN model ranges from thousands to millions. Due to the long execution time, it is impractical to explore the impact of all weight bits perturbations on a DNN model. 
Therefore, to find the most vulnerable bits precisely and effectively, we introduce an iterative method based on a gradient ranking and iterative search. In each iteration, it finds the top $n$ most vulnerable wights in the $l$-th layer (i.e., $b_l$) through gradient ranking.
To do so, we compute the gradient of adversarial robustness function with respect to weight bits $\bm{b}$ and rank them in descending order. Then, considering Equation \ref{mask}, we apply bit flip as:
\begin{equation}
    \bm{\hat{b_l}} = \bm{b_l} \oplus \bm{m}
\end{equation}
where $\bm{\hat{b_l}}$ shows the flipped bit.

\begin{table*}[tb!]
\begin{center}
\caption {Overall accuracy (for clean inputs) and robustness (for crafted inputs) results of the original and \textbf{protected} DNN models before and after our bit-flip attack.}
\vspace{-0.05in}   
\begin{tabular}{|l|l|c|c|c|c|c|}
\hline
\multirow{2}{*}{\begin{tabular}[c]{@{}l@{}} DNN Model\end{tabular}} & \multirow{2}{*}{Dataset} & \multicolumn{2}{c|}{Accuracy (Clean Inputs)} & \multicolumn{3}{c|}{Robustness (Crafted Inputs)} \\ \cline{3-7} 
 &  & Protected Model & \textbf{Our Attack} &  Original Model & Protected Model  & \textbf{Our Attack (Drop \%)} \\ \hline \hline
LeNet-3 & MNIST & 0.9912 & 0.9844 & 0.154 & 0.591 & 0.237 (59.88\%)  \\ \hline
FCN-2 & MNIST & 0.8935 & 0.8895 & 0.081 & 0.716 & 0.194 (72.90\%) \\ \hline
LeNet-5 & CIFAR-10 & 0.7983 & 0.789 & 0.163 & 0.504 & 0.201 (60.11\%) \\ \hline
AlexNet & CIFAR-10 & 0.7263 & 0.720 & 0.096 & 0.683 & 0.196 (71.30\%) \\ \hline
\end{tabular}
\label{tab:MainResults}
\end{center}
\vspace{-0.1in}
\end{table*}

Afterwards, it records the adversarial robustness of each layer and selects the most vulnerable bits in the entire DNN model among the candidate bits of all layers. 
This iterative algorithm terminates when the robustness drop becomes sufficiently close to an expected value or the accuracy drop becomes higher than the attacker-defined constraint ($\delta$ in Equation \ref{optimization}).

\section{Experimental Results}
\label{sec:results}


\subsection{Experimental Setup}\label{subsec:setup}

\noindent \textbf{Implementation Framework:}
Our proposed attack is implemented in Python on top of the PyTorch framework \cite{PyTorch}. We perform all our experiments on a Nvidia Titan V GPU. 

\noindent \textbf{Dataset and DNN Models:}
We choose two widely used visual datasets for image classification, including MNIST and CIFAR-10.
For MNIST, we use the well-known LeNet-3 and FCN-2  network models. For CIFAR-10, we use  the well-known LeNet-5 and AlexNet network models. 
For all DNN models, we use 8-bit fixed-point numbers for the DNN weight parameters.

\noindent \textbf{Protection Algorithm:}
In our experiments, we use two types of DNN models: the original models and the protected ones.
To generate adversarial examples, we leverage the widely used DeepFool adversarial attack tool \cite{moosavi2016deepfool}. 
To protect DNNs against conventional adversarial example attacks, we use the state-of-the-art defense algorithm called TRADES \cite{zhang2019theoretically}, 
which retrains models via a mixture of clean inputs and adversarial examples with additional training epochs. 

\subsection{Overall Accuracy and Robustness Results}\label{subsec:overall-results}

Table~\ref{tab:MainResults} summarizes the overall accuracy and robustness results for the protected DNN models, both before and after our bit-flip attack. 
First, the accuracy loss of all models under our attack is within 1\%, which makes users hard to notice for clean inputs. Second, under our bit-flip attack, the robustness drops significantly by 59.9\% to 72.9\% which indicates that it could circumvent the TRADES defense using adversarial examples. 
To better illustrate the impact of the robustness drop, we also measure the robustness of the original DNN models (without any protection), which is also shown in Table~\ref{tab:MainResults}.
The robustness metric of the protected DNNs under our bit-flip attack becomes low, which suggests that we successfully compromised the robustness of protected DNNs (with negligible accuracy loss).

\subsection{Number of Perturbed DNN Weight Bits} \label{subsec:pertrubed-weights}

Table~\ref{tab:bitflips} summarizes the number of bit-flips that is required for each DNN model to achieve our attack, which is a very small amount that ranges from 30 to 101 bit-flips.
Generally, we require less than 0.006\% of the weight bits (note each weight has 8 bits) to be perturbed. This indicates that our attack could be easily and effectively deployed in the hardware.


\begin{table}[h!]
\begin{center}
\caption{Number of bit flips in our attack. Note that each weight has 8 bits in our DNN models.}
\label{tab:bitflips}
\vspace{-0.05in}    
\begin{tabular}{|l|c|c|c|}
        \hline
         DNN Model & \#Total Weights&\#Bit Flips & Flip Percentage
         \\ \hline\hline
         LeNet-3& 61,706 &30&0.0060\%\\ \hline
         FCN-2&589,160 &49&0.0010\% \\ \hline
         LeNet-5&657,080 &71&0.0013\% \\ \hline
         AlexNet&2,472,266 &101&0.0005\%\\ \hline
         \end{tabular}
    \vspace{-0.1in}
\end{center}
\end{table}

\subsection{Case Study of Our Attack on Protected DNNs} \label{subsec:comp-prior}
To better illustrate the effectiveness of our stealthy attack on protected DNNs, we conduct a case study of our attack over the protected LeNet-3 on the MNIST dataset and protected LeNet-5 on CIFAR-10 dataset.

\begin{figure}[!tb]
	\begin{center}
	\includegraphics[width=1.7in]{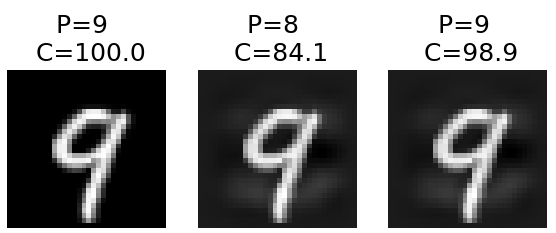}
    \includegraphics[width=1.7in]{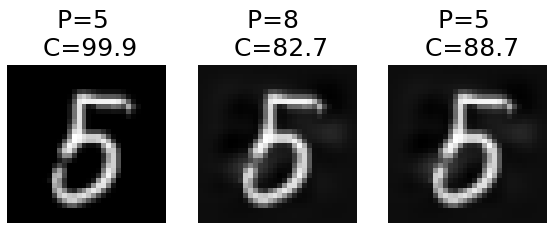}
    \includegraphics[width=1.7in]{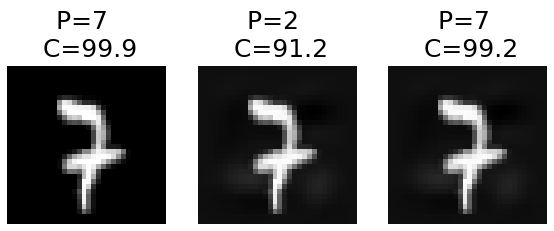}
    \includegraphics[width=1.7in]{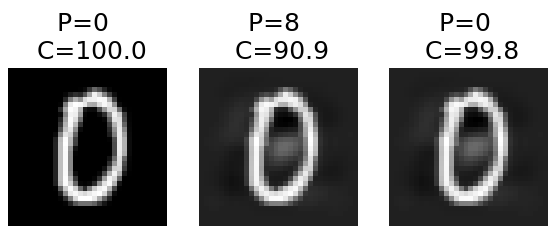}
    \includegraphics[width=1.7in]{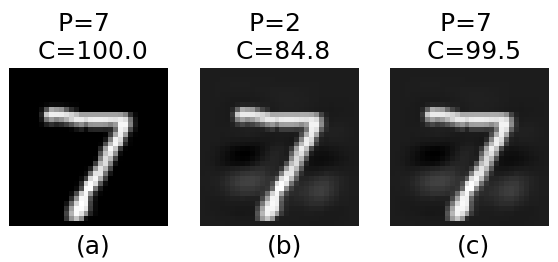}
    \includegraphics[width=1.7in]{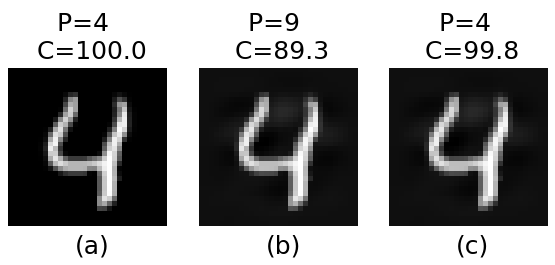}
	\end{center}
	\vspace{-0.1in}
	\caption{Comparison of prediction (P=..) class with the highest confidence (C=..\%) by the \textbf{protected} LeNet-3 on MNIST dataset: (a) clean inputs, our attack; (b) crafted inputs, our attack; (c) crafted inputs, no (hardware parameter) attack.}
	\label{fig:comp-protected}
	\vspace{-0.1in}
\end{figure}

Figure~\ref{fig:comp-protected} shows some example classification results---i.e., the classification class with the highest confidence---for both clean and crafted inputs. 
The subfigures (a) confirm that the protected DNN still correctly classifies the clean images under our attack. 
The subfigures (b) confirm that the protected DNN misclassifies the crafted images under our attack. These two points further confirm that we have successfully achieved the stealthy attack on protected DNNs.
The subfigures (c) confirm that the protected DNN still correctly classifies the crafted images without our attack on hardware parameters; that is, prior adversarial input attacks do not work for protected DNNs. 

\begin{figure}[!tb]
	\begin{center}
    \includegraphics[width=0.65in]{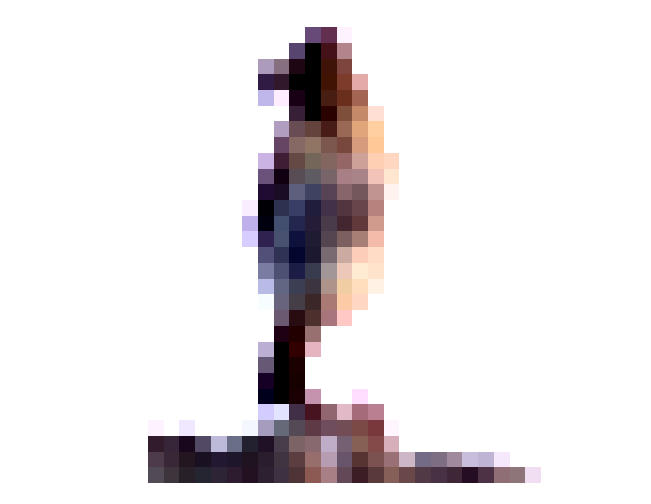}\hspace{0.03em}
    \includegraphics[width=0.65in]{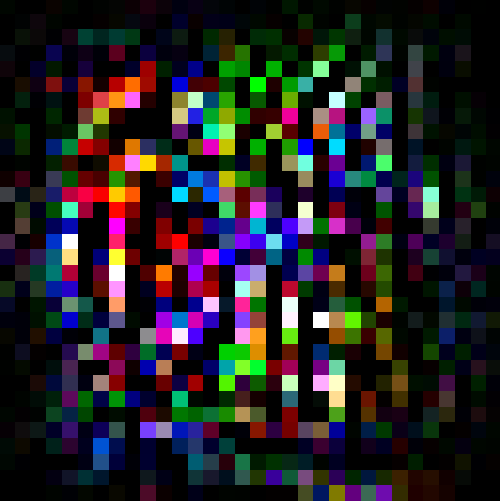}\hspace{0.03em}
    \includegraphics[width=0.65in]{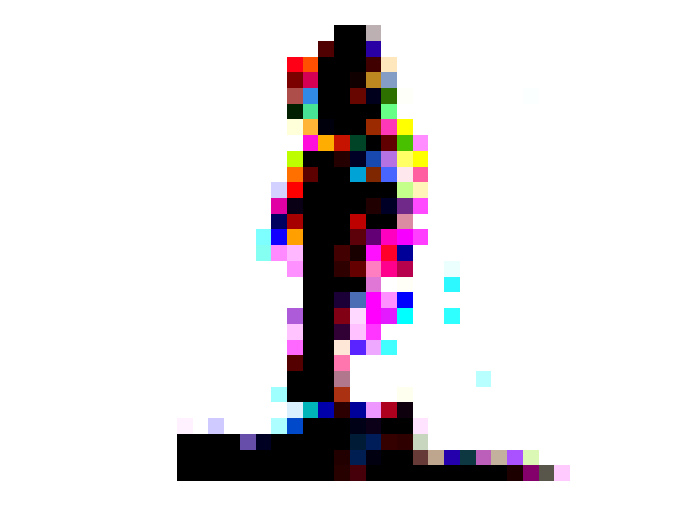}\hspace{0.03em}
    \includegraphics[width=0.65in]{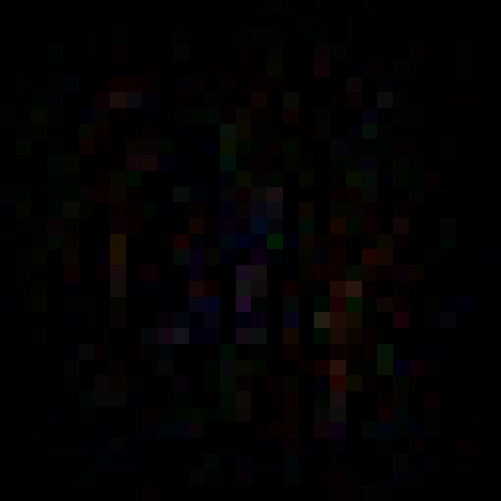}\hspace{0.03em}
    \includegraphics[width=0.65in]{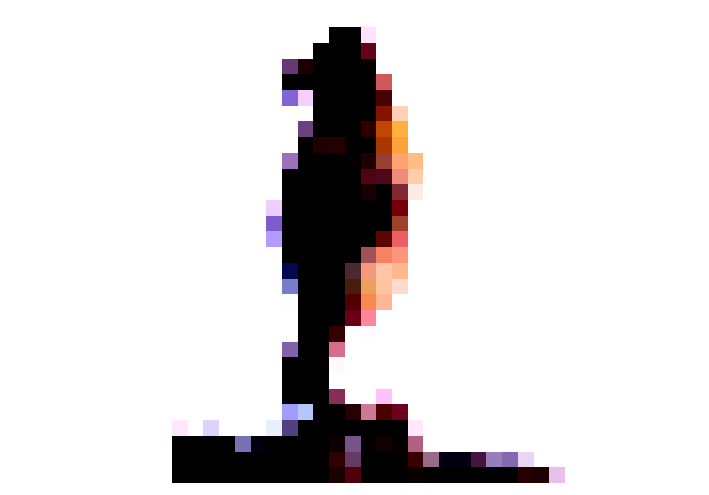}\hspace{0.03em}
    \\\vspace{0.2em}
    \subcaptionbox{\label{perturb:a}}{\includegraphics[width=0.65in]{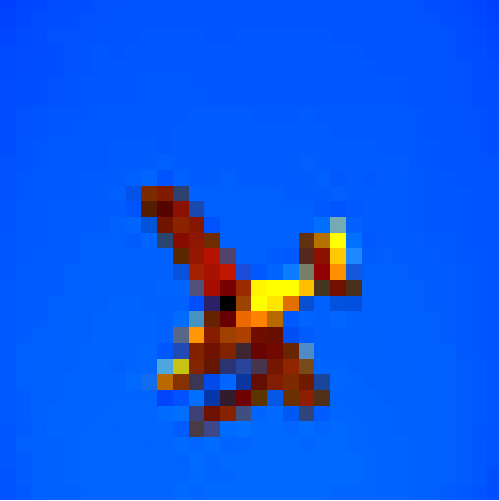}\hspace{0.03em}}
    \subcaptionbox{\label{perturb:b}}{\includegraphics[width=0.65in]{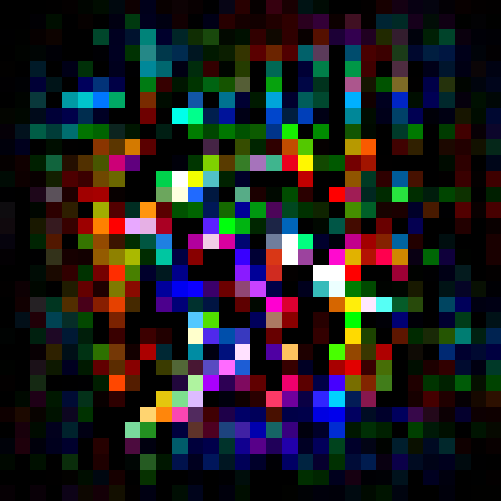}\hspace{0.03em}}
    \subcaptionbox{\label{perturb:c}}{\includegraphics[width=0.65in]{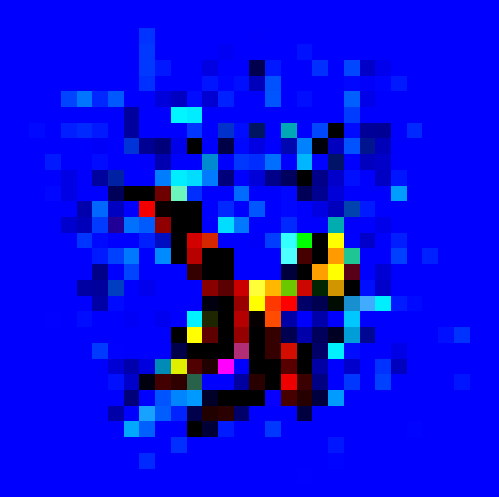}\hspace{0.03em}}
    \subcaptionbox{\label{perturb:d}}{\includegraphics[width=0.65in]{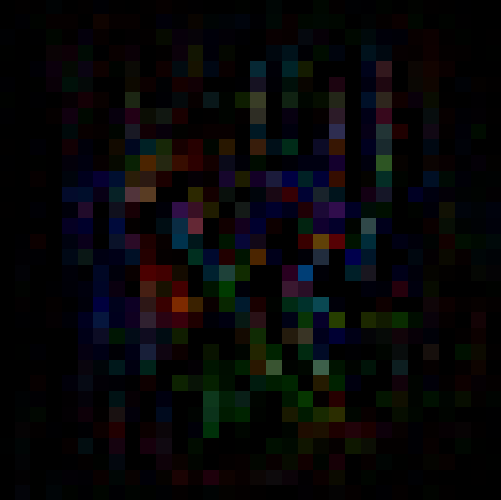}\hspace{0.03em}}
    \subcaptionbox{\label{perturb:e}}{\includegraphics[width=0.65in]{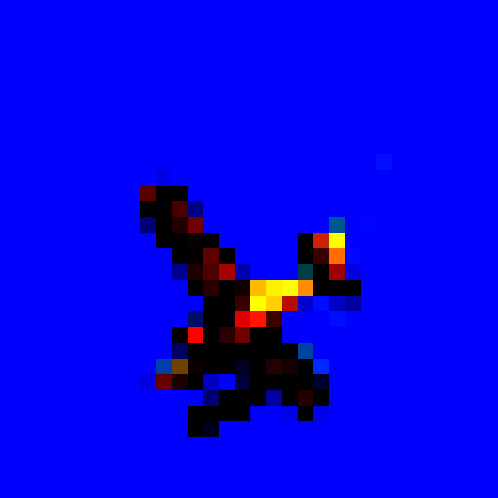}\hspace{0.03em}}
	\caption{Comparison of the input perturbation vector required to fool the \textbf{protected} LeNet-5 on CIFAR-10 dataset: (a) clean inputs; (b) perturbation vectors required by DeepFool~\cite{moosavi2016deepfool}; (c) crafted inputs for DeepFool~\cite{moosavi2016deepfool}; (d) perturbation vectors required by our attack; (e) crafted inputs for our attack.}
	\label{fig:comp-unprotected}
 	\vspace{-0.2in}
	\end{center}
\end{figure}

Figure~\ref{fig:comp-unprotected} shows some example input perturbations required to fool the protected LeNet-5 on CIFAR-10 dataset. Figure~\ref{fig:comp-unprotected}(a) shows the clean images; (b) and (c) show the input perturbation vectors and crafted images required by DeepFool~\cite{moosavi2016deepfool} to fool protected LeNet-5; (d) and (e) show the input perturbation vectors and crafted images required by our bit-flip attack to fool protected LeNet-5. It can be easily seen by human eyes that our attack requires much less perturbations to the input images in order to fool a protected DNN.


\section{Conclusion and Discussion}
\label{sec:concl}

Continuing their great success in many areas, DNNs are being deployed in critical systems such as personal identity recognition systems and autonomous vehicles. This creates great security concerns about the DNN deployment. In the recent years, researchers have already investigated various types of adversarial example attacks on DNNs and proposed algorithmic countermeasures for adversarial examples. In this paper, we have proposed a new type of stealthy bit-flip attack on protected DNNs to compromise their robustness while reserving their accuracy, by attacking the DNN weight parameters in the hardware. We mathematically formulate this stealthy attack as an optimization problem and introduce a gradient-based algorithm to efficiently find the most vulnerable weight bits. Experimental results demonstrate that the robustness of protected DNNs can significantly decrease under our adversarial attack with a small number of bit-flips, while there is negligible accuracy loss for clean inputs. Our attack on TRADES \cite{zhang2019theoretically} protection-based models can decrease the robustness value by $59.9\%$ to $72.9\%$.

Our proposed stealthy attack opens new opportunities regarding the attack and defense of DNNs with an emphasis on software-hardware co-design.  
In future work, we plan to investigate corresponding defenses for this new type of attack. For example, one may attempt to reconstruct DNN weights in a way such that the change in a weight value diffuses to its neighbor weights and hence removing the stealthy attribute.

\section*{Acknowledgements}
We acknowledge the support from Government of Canada Technology Demonstration Program and MDA Systems Ltd; NSERC Discovery Grant RGPIN341516, RGPIN-2019-04613, DGECR-2019-00120, Alliance Grant ALLRP-552042-2020, COHESA (NETGP485577-15), CWSE PDF (470957); CFI John R. Evans Leaders Fund; Simon Fraser University New Faculty Start-up Grant.

\bibliographystyle{IEEEtran}
\bibliography{references}

\begin{thebibliography}{10}
\providecommand{\url}[1]{#1}
\csname url@samestyle\endcsname
\providecommand{\newblock}{\relax}
\providecommand{\bibinfo}[2]{#2}
\providecommand{\BIBentrySTDinterwordspacing}{\spaceskip=0pt\relax}
\providecommand{\BIBentryALTinterwordstretchfactor}{4}
\providecommand{\BIBentryALTinterwordspacing}{\spaceskip=\fontdimen2\font plus
\BIBentryALTinterwordstretchfactor\fontdimen3\font minus
  \fontdimen4\font\relax}
\providecommand{\BIBforeignlanguage}[2]{{%
\expandafter\ifx\csname l@#1\endcsname\relax
\typeout{** WARNING: IEEEtran.bst: No hyphenation pattern has been}%
\typeout{** loaded for the language `#1'. Using the pattern for}%
\typeout{** the default language instead.}%
\else
\language=\csname l@#1\endcsname
\fi
#2}}
\providecommand{\BIBdecl}{\relax}
\BIBdecl

\bibitem{shafique2020robust}
M.~Shafique, M.~Naseer, T.~Theocharides, C.~Kyrkou, O.~Mutlu, L.~Orosa, and
  J.~Choi, ``Robust machine learning systems: Challenges, current trends,
  perspectives, and the road ahead,'' \emph{IEEE Design \& Test}, vol.~37,
  no.~2, pp. 30--57, 2020.

\bibitem{szegedy2013intriguing}
C.~Szegedy, W.~Zaremba, I.~Sutskever, J.~Bruna, D.~Erhan, I.~Goodfellow, and
  R.~Fergus, ``Intriguing properties of neural networks,'' \emph{arXiv preprint
  arXiv:1312.6199}, 2013.

\bibitem{goodfellow2014explaining}
I.~J. Goodfellow, J.~Shlens, and C.~Szegedy, ``Explaining and harnessing
  adversarial examples,'' \emph{arXiv preprint arXiv:1412.6572}, 2014.

\bibitem{moosavi2016deepfool}
S.-M. Moosavi-Dezfooli, A.~Fawzi, and P.~Frossard, ``Deepfool: a simple and
  accurate method to fool deep neural networks,'' in \emph{Proceedings of the
  IEEE conference on computer vision and pattern recognition}, 2016, pp.
  2574--2582.

\bibitem{carlini2017towards}
N.~Carlini and D.~Wagner, ``Towards evaluating the robustness of neural
  networks,'' in \emph{2017 ieee symposium on security and privacy (sp)}.\hskip
  1em plus 0.5em minus 0.4em\relax IEEE, 2017, pp. 39--57.

\bibitem{moosavi2017universal}
S.-M. Moosavi-Dezfooli, A.~Fawzi, O.~Fawzi, and P.~Frossard, ``Universal
  adversarial perturbations,'' in \emph{Proceedings of the IEEE conference on
  computer vision and pattern recognition}, 2017, pp. 1765--1773.

\bibitem{hayes2018learning}
J.~Hayes and G.~Danezis, ``Learning universal adversarial perturbations with
  generative models,'' in \emph{2018 IEEE Security and Privacy Workshops
  (SPW)}.\hskip 1em plus 0.5em minus 0.4em\relax IEEE, 2018, pp. 43--49.

\bibitem{yuan2019adversarial}
X.~Yuan, P.~He, Q.~Zhu, and X.~Li, ``Adversarial examples: Attacks and defenses
  for deep learning,'' \emph{IEEE transactions on neural networks and learning
  systems}, vol.~30, no.~9, pp. 2805--2824, 2019.

\bibitem{chaubey2020universal}
A.~Chaubey, N.~Agrawal, K.~Barnwal, K.~K. Guliani, and P.~Mehta, ``Universal
  adversarial perturbations: A survey,'' \emph{arXiv preprint
  arXiv:2005.08087}, 2020.

\bibitem{chakraborty2018adversarial}
A.~Chakraborty, M.~Alam, V.~Dey, A.~Chattopadhyay, and D.~Mukhopadhyay,
  ``Adversarial attacks and defences: A survey,'' \emph{arXiv preprint
  arXiv:1810.00069}, 2018.

\bibitem{sinha2017certifying}
A.~Sinha, H.~Namkoong, R.~Volpi, and J.~Duchi, ``Certifying some distributional
  robustness with principled adversarial training,'' \emph{arXiv preprint
  arXiv:1710.10571}, 2017.

\bibitem{madry2017towards}
A.~Madry, A.~Makelov, L.~Schmidt, D.~Tsipras, and A.~Vladu, ``Towards deep
  learning models resistant to adversarial attacks,'' \emph{arXiv preprint
  arXiv:1706.06083}, 2017.

\bibitem{zhang2019theoretically}
H.~Zhang, Y.~Yu, J.~Jiao, E.~Xing, L.~El~Ghaoui, and M.~Jordan, ``Theoretically
  principled trade-off between robustness and accuracy,'' in
  \emph{International Conference on Machine Learning}.\hskip 1em plus 0.5em
  minus 0.4em\relax PMLR, 2019, pp. 7472--7482.

\bibitem{wong2020fast}
E.~Wong, L.~Rice, and J.~Z. Kolter, ``Fast is better than free: Revisiting
  adversarial training,'' \emph{arXiv preprint arXiv:2001.03994}, 2020.

\bibitem{liu2017fault}
Y.~Liu, L.~Wei, B.~Luo, and Q.~Xu, ``Fault injection attack on deep neural
  network,'' in \emph{2017 IEEE/ACM International Conference on Computer-Aided
  Design (ICCAD)}.\hskip 1em plus 0.5em minus 0.4em\relax IEEE, 2017, pp.
  131--138.

\bibitem{breier2018practical}
J.~Breier, X.~Hou, D.~Jap, L.~Ma, S.~Bhasin, and Y.~Liu, ``Practical fault
  attack on deep neural networks,'' in \emph{Proceedings of the 2018 ACM SIGSAC
  Conference on Computer and Communications Security}, 2018, pp. 2204--2206.

\bibitem{rakin2019bit}
A.~S. Rakin, Z.~He, and D.~Fan, ``Bit-flip attack: Crushing neural network with
  progressive bit search,'' in \emph{Proceedings of the IEEE International
  Conference on Computer Vision}, 2019, pp. 1211--1220.

\bibitem{yao2020deephammer}
F.~Yao, A.~S. Rakin, and D.~Fan, ``Deephammer: Depleting the intelligence of
  deep neural networks through targeted chain of bit flips,'' \emph{arXiv
  preprint arXiv:2003.13746}, 2020.

\bibitem{dumont2021overview}
M.~Dumont, P.-A. Moellic, R.~Viera, J.-M. Dutertre, and R.~Bernhard, ``An
  overview of laser injection against embedded neural network models,''
  \emph{arXiv preprint arXiv:2105.01403}, 2021.

\bibitem{zhao2019fault}
P.~Zhao, S.~Wang, C.~Gongye, Y.~Wang, Y.~Fei, and X.~Lin, ``Fault sneaking
  attack: A stealthy framework for misleading deep neural networks,'' in
  \emph{2019 56th ACM/IEEE Design Automation Conference (DAC)}.\hskip 1em plus
  0.5em minus 0.4em\relax IEEE, 2019, pp. 1--6.

\bibitem{rakin2020t}
A.~S. Rakin, Z.~He, J.~Li, F.~Yao, C.~Chakrabarti, and D.~Fan, ``T-bfa:
  Targeted bit-flip adversarial weight attack,'' \emph{arXiv preprint
  arXiv:2007.12336}, 2020.

\bibitem{bai2021targeted}
J.~Bai, B.~Wu, Y.~Zhang, Y.~Li, Z.~Li, and S.-T. Xia, ``Targeted attack against
  deep neural networks via flipping limited weight bits,'' \emph{arXiv preprint
  arXiv:2102.10496}, 2021.

\bibitem{liu2017trojaning}
Y.~Liu, S.~Ma, Y.~Aafer, W.-C. Lee, J.~Zhai, W.~Wang, and X.~Zhang, ``Trojaning
  attack on neural networks,'' \emph{NDSS 2018}.

\bibitem{li2018hu}
W.~Li, J.~Yu, X.~Ning, P.~Wang, Q.~Wei, Y.~Wang, and H.~Yang, ``Hu-fu: Hardware
  and software collaborative attack framework against neural networks,'' in
  \emph{2018 IEEE Computer Society Annual Symposium on VLSI (ISVLSI)}.\hskip
  1em plus 0.5em minus 0.4em\relax IEEE, 2018, pp. 482--487.

\bibitem{clements2018hardware}
J.~Clements and Y.~Lao, ``Hardware trojan attacks on neural networks,''
  \emph{arXiv preprint arXiv:1806.05768}, 2018.

\bibitem{zhao2019memory}
Y.~Zhao, X.~Hu, S.~Li, J.~Ye, L.~Deng, Y.~Ji, J.~Xu, D.~Wu, and Y.~Xie,
  ``Memory trojan attack on neural network accelerators,'' in \emph{2019
  Design, Automation \& Test in Europe Conference \& Exhibition (DATE)}.\hskip
  1em plus 0.5em minus 0.4em\relax IEEE, 2019, pp. 1415--1420.

\bibitem{rakin2020tbt}
A.~S. Rakin, Z.~He, and D.~Fan, ``Tbt: Targeted neural network attack with bit
  trojan,'' in \emph{Proceedings of the IEEE/CVF Conference on Computer Vision
  and Pattern Recognition}, 2020, pp. 13\,198--13\,207.

\bibitem{kim2014flipping}
Y.~Kim, R.~Daly, J.~Kim, C.~Fallin, J.~H. Lee, D.~Lee, C.~Wilkerson, K.~Lai,
  and O.~Mutlu, ``Flipping bits in memory without accessing them: An
  experimental study of dram disturbance errors,'' \emph{ACM SIGARCH Computer
  Architecture News}, vol.~42, no.~3, pp. 361--372, 2014.

\bibitem{mopuri2017fast}
{Mopuri et al.}, ``Fast feature fool: A data independent approach to universal
  adversarial perturbations,'' \emph{arXiv preprint arXiv:1707.05572}, 2017.

\bibitem{mopuri2018generalizable}
K.~R. Mopuri, A.~Ganeshan, and R.~V. Babu, ``Generalizable data-free objective
  for crafting universal adversarial perturbations,'' \emph{IEEE transactions
  on pattern analysis and machine intelligence}, vol.~41, no.~10, pp.
  2452--2465, 2018.

\bibitem{bai2021recent}
T.~Bai, J.~Luo, J.~Zhao, and B.~Wen, ``Recent advances in adversarial training
  for adversarial robustness,'' \emph{arXiv preprint arXiv:2102.01356}, 2021.

\bibitem{modelzoosite}
\BIBentryALTinterwordspacing
Model zoo: Discover open source deep learning code and pretrained models.
  [Online]. Available: \url{https://modelzoo.co}
\BIBentrySTDinterwordspacing

\bibitem{gongye2020reverse}
C.~Gongye, Y.~Fei, and T.~Wahl, ``Reverse-engineering deep neural networks
  using floating-point timing side-channels,'' in \emph{2020 57th ACM/IEEE
  Design Automation Conference (DAC)}.\hskip 1em plus 0.5em minus 0.4em\relax
  IEEE, 2020, pp. 1--6.

\bibitem{wei2020leaky}
J.~Wei, Y.~Zhang, Z.~Zhou, Z.~Li, and M.~A. Al~Faruque, ``Leaky dnn: Stealing
  deep-learning model secret with gpu context-switching side-channel,'' in
  \emph{2020 50th Annual IEEE/IFIP International Conference on Dependable
  Systems and Networks (DSN)}.\hskip 1em plus 0.5em minus 0.4em\relax IEEE,
  2020, pp. 125--137.

\bibitem{xiang2020open}
Y.~Xiang, Z.~Chen, Z.~Chen, Z.~Fang, H.~Hao, J.~Chen, Y.~Liu, Z.~Wu, Q.~Xuan,
  and X.~Yang, ``Open dnn box by power side-channel attack,'' \emph{IEEE
  Transactions on Circuits and Systems II: Express Briefs}, vol.~67, no.~11,
  pp. 2717--2721, 2020.

\bibitem{kurakin2016adversarial}
A.~Kurakin, I.~Goodfellow, S.~Bengio \emph{et~al.}, ``Adversarial examples in
  the physical world,'' 2016.

\bibitem{sharif2016accessorize}
M.~Sharif, S.~Bhagavatula, L.~Bauer, and M.~K. Reiter, ``Accessorize to a
  crime: Real and stealthy attacks on state-of-the-art face recognition,'' in
  \emph{Proceedings of the 2016 acm sigsac conference on computer and
  communications security}, 2016, pp. 1528--1540.

\bibitem{tatar2018defeating}
A.~Tatar, C.~Giuffrida, H.~Bos, and K.~Razavi, ``Defeating software mitigations
  against rowhammer: a surgical precision hammer,'' in \emph{International
  Symposium on Research in Attacks, Intrusions, and Defenses}.\hskip 1em plus
  0.5em minus 0.4em\relax Springer, 2018, pp. 47--66.

\bibitem{goodfellow2016deep}
I.~Goodfellow, Y.~Bengio, A.~Courville, and Y.~Bengio, \emph{Deep
  learning}.\hskip 1em plus 0.5em minus 0.4em\relax MIT press Cambridge, 2016,
  vol.~1, no.~2.

\bibitem{PyTorch}
{Paszke et al.}, ``Pytorch: An imperative style, high-performance deep learning
  library,'' in \emph{NeurIPS 2019}, pp. 8024--8035.

\end{thebibliography}

\end{document}